\documentclass[a4paper,11pt]{article}
\pdfoutput=1
\usepackage{jheppub}
\usepackage[utf8]{inputenc}

\usepackage{tikz}
\usetikzlibrary{positioning}

\renewcommand{\d}{\mathrm{d}}

\tikzset{
smallnode/.style={circle, draw, very thick, minimum size=2mm},
smallcircle/.style={circle, fill, scale=0.5},
roundnode/.style={circle, draw, very thick, minimum size=10mm},
squarednode/.style={rectangle, draw, very thick, minimum size=10mm},
roundSO/.style={circle, draw, fill=gray!100, very thick, minimum size=10mm},
roundUSp/.style={circle, draw, fill=black, very thick, minimum size=10mm, text=white},
squareSO/.style={rectangle, draw, fill=gray!100, very thick, minimum size=10mm},
squareUSp/.style={rectangle, draw, fill=black, very thick, minimum size=10mm, text=white},
c1/.style={circle, draw, very thick, minimum size=1mm},
c2/.style={circle, draw, fill=black!100, very thick, minimum size=1mm}
}
\tikzset{every loop/.style={}}

\title{\boldmath Non-Connected Gauge Groups and the Plethystic Program}

\author{Antoine Bourget and Alessandro Pini}

\affiliation{Department of Physics, Universidad de Oviedo,\\ Avenida Calvo Sotelo 18, 33007 Oviedo, Spain}
\emailAdd{bourgetantoine@uniovi.es}
\emailAdd{pinialessandro@uniovi.es}

\abstract{We present in the context of supersymmetric gauge theories an extension of the Weyl integration formula, first discovered by Robert Wendt \cite{wendt2001weyl}, which applies to a class of non-connected Lie groups. This allows to count in a systematic way gauge-invariant chiral operators for these non-connected gauge groups. Applying this technique to $\mathrm{O}(n)$, we obtain, via the ADHM construction, the Hilbert series for certain instanton moduli spaces. We validate our general method and check our results via a Coulomb branch computation, using three-dimensional mirror symmetry. }

\begin{document} 
\maketitle
\flushbottom

\section{Introduction}

The study of supersymmetric gauge theories in various numbers of dimensions is of great interest since we can compute a wealth of observables in an exact way. In particular, the counting of chiral gauge invariant operators is a central task whose completion allows to understand various aspects of the QFT under consideration. This procedure has been successfully addressed in the context of the so-called Plethystic Program \cite{Benvenuti:2006qr}, where a crucial role is played by the Hilbert series, i.e. a generating function that counts the chiral operators present in the theory according to their dimension and other quantum numbers (see \cite{Cremonesi:2017jrk} for a recent review). Moreover Higgs and Coulomb branch Hilbert series of many three- and four-dimensional theories with extended supersymmetry have been computed and analyzed extensively \cite{Benvenuti:2010pq,Hanany:2012dm,Cremonesi:2013lqa,Cremonesi:2014xha,Cremonesi:2015dja}. 

The cases in which the gauge group of the theory is connected have been thoroughly studied. On the other hand, the realm of gauge theories with non-connected gauge groups has received much less attention from this perspective, although four-dimensional $\mathcal{N}=2$ theories of this type have been studied recently \cite{Argyres:2016yzz}. 
In addition of being an inescapable step towards a full understanding of gauge quantum field theories, some theories with non-connected gauge groups are in fact needed in a variety of common situations. For instance, the moduli space of $k$ $\mathrm{Sp}(N)$ instantons on $\mathbb{R}^4$ is related through the Atiyah-Hitchin-Drinfeld-Manin (ADHM) construction \cite{adhm} to a gauge theory with a full orthogonal gauge group $\mathrm{O}(k)$ \cite{Benvenuti:2010pq,Hanany:2012dm}. In order to characterize this moduli space of instantons, it is therefore necessary to deal with the disconnected gauge group $\mathrm{O}(k)$, thus following a road paved with unexpected technical impediments. 

In this article, we outline a general procedure first presented in \cite{wendt2001weyl} that allows to count gauge invariant operators, through integration over the gauge group of the theory using an extension of the Weyl integration formula, that applies to a class of non-connected gauge Lie groups called the \emph{principal extensions}. To the best of our knowledge, this formula has never been applied to quantum field theory before. In the case of the $D_n$ simple Lie algebra, the principal extension is just $\mathrm{O}(2n)$, as we show in section \ref{subsection:O(2n)}. This means that the integration formula can be applied to problems of gauge invariant operator counting in theories with $\mathrm{O}(2n)$ gauge groups. A particularly interesting example of such a computation is the Hilbert series of the moduli space of $\mathrm{Sp}(N)$ instantons, as mentioned previously. In the case of Lie algebras of type $A$ and $E$, the principal extension produces more exotic candidates for defining gauge theories. These exotic gauge theories will be defined and characterized elsewhere \cite{Bourget:2018ond}. 

This article is organized as follows. In Section \ref{sectionIntegration}, we review the construction of the principal extensions and write down the general integration formula on a non-connected component of a group. Then we focus on the case of $\mathrm{O}(n)$ groups, and summarize the integration procedure in equation (\ref{OnIntegration}). Then in section \ref{sectionModuliSpace}, we use these results to compute the Hilbert series for the moduli space of $k$ $\mathrm{Sp}(N)$ instantons, giving explicit results for small values of $k$ and $N$. This fills a gap in the literature, and also sheds new light on previous computations. We then provide two checks of the validity of our results: 
\begin{itemize}
    \item Using a letter counting argument, combined with an analysis of the gauge-invariance conditions, we can reproduce to arbitrary order the series expansion of the Hilbert series; 
    \item Combining three-dimensional mirror symmetry \cite{Intriligator:1996ex} applied to the ADHM quiver and the monopole formula \cite{Cremonesi:2013lqa}, our results can be cross-checked through the computation of the Coulomb branch Hilbert series of the mirror theory. Indeed we find agreement in all cases. 
\end{itemize}
The text is completed by several appendices containing some background in group theory, the proof of some properties used in Section \ref{sectionIntegration}, and the results of some of our computations.

\section{\texorpdfstring{The Generalized Weyl Formula}{}}
\label{sectionIntegration}

In this section, we first review the well-known proof of the Weyl integration formula for a connected Lie group. This serves as a reminder and as a warm-up for the next subsection where we consider the non-connected case in the framework of principal extensions, which will be defined there. Then we show how this fairly abstract construction can be applied to the well-known example of $\mathrm{O}(k)$ groups, and how it can be visualized via brane constructions in string theory. The readers who are more interested in physical considerations can skip all of this section, taking for granted the formula (\ref{OnIntegration}) that is used in subsequent developments. 

\subsection{The Weyl formula for connected groups}
\label{sectionWeyl}

Let $G$ be a rank $r$ compact \emph{connected} and simply connected semisimple Lie group with invariant and normalized Haar measure\footnote{We use the generic notation $\d \eta_G$ for the Haar measure of the group $G$, since the letter $\eta$ is the Greek equivalent of the letter $h$. We will keep the notations $\d \tilde{\mu}_G$ and $\d \mu_G$ for the Haar measure on a maximal torus of $G$ which in addition includes the Jacobian for the change of variables. See equations (\ref{defmutilde}) and (\ref{defmu}).} $\d \eta_G$, and let $T$ be a maximal torus of $G$, with invariant and normalized Haar measure $\d \eta_T$. We call $\mathfrak{g}$ and $\mathfrak{t}$ their respective Lie algebras. We choose fugacities $z_1 , \dots , z_r$ to parameterize $T$ so that this measure is simply 
\begin{equation}
   \int_T  \d \eta_T = \prod\limits_{l=1}^r \oint_{|z_l|=1} \frac{\d z_l}{z_l} \, .  
\end{equation}
Let $W$ be the Weyl group associated to $T$. Consider an integrable function $f$ that is constant on conjugacy classes, for which we want to compute 
\begin{equation}
     \int_G  f(X) \d \eta_G(X)  \, . 
\end{equation}
The idea of the Weyl formula is that since in a connected group any $X \in G$ is conjugate to an element $z \in T$ (see for instance Theorem 4.36 in \cite{knapp2013lie}), we can perform the change of variables $X = yzy^{-1}$ with $y \in G/T$ and $z \in T$. Let us define accordingly the map 
\begin{eqnarray}
\label{psimap}
     \psi &:& G/T \times T \rightarrow G \\
     & & (y,z) \mapsto yzy^{-1} \, .  \nonumber
\end{eqnarray}
One can show that this map is\footnote{Excepted on a measure zero subset of $G/T \times T$. Here and in the following, $|W|$ is the cardinality of $W$. } $|W|$-to-one, and that the associated Jacobian for the change of coordinates $X = \psi (y,z)$ is (see Appendix \ref{appendixJacobian} for the details of the computation)
\begin{equation}
\label{detDiff1}
     \det (\d \psi)_{(y,z)} = \det \left( \mathrm{Ad}(z^{-1})-1 \right) |_{\mathfrak{t}^{\perp}} = \prod\limits_{\alpha \in \Delta (G)} (1-z^{-\alpha}) \, ,  
\end{equation}
where $\mathfrak{t}^{\perp}$ is the complement of $\mathfrak{t}$ in $\mathfrak{g}$, and $\Delta (G)$ is the set of roots of $G$. Putting everything together, we obtain the Weyl integration formula,  
\begin{equation}
\label{WeylFormula1}
     \int_G f(X) \d \eta_G(X)  = \frac{1}{|W|} \int_T  f(z) \prod\limits_{\alpha \in \Delta(G)} (1-z^{-\alpha}) \d \eta_T (z) \, . 
\end{equation}

Let us now introduce two related but different measures on the maximal torus, 
\begin{equation}
\label{defmutilde}
     \d \tilde{\mu}_G (z) = \frac{1}{|W|} \prod\limits_{\alpha \in \Delta(G)} (1-z^{-\alpha}) \d \eta_T (z) \,  ,
\end{equation}
and 
\begin{equation}
\label{defmu}
     \d \mu_G (z) = \prod\limits_{\alpha \in \Delta^+(G)} (1-z^{-\alpha}) \d \eta_T (z) \, . 
\end{equation}
As we prove in Appendix \ref{appendixMeasures}, if the function $f(z)$ is invariant under the Weyl group, then we have 
\begin{equation}
\label{equalityMeasures}
     \int_T f(z) \d \mu_G (z) = \int_T f(z) \d \tilde{\mu}_G (z)\, . 
\end{equation}
This applies in particular to the character of any finite representation of $G$. The measure $\d \mu_G (z)$ involves a polynomial with $\frac{1}{2}(\mathrm{dim} G - r)$ factors, twice less than in the polynomial involved in $\d \tilde{\mu}_G (z)$, so we will use the former. Combining (\ref{WeylFormula1}) and (\ref{equalityMeasures}), one obtains the Weyl integration formula in the form that will be most useful to us, 
\begin{equation}
\label{WeylFormula2}
    \int_G f(X) \d \eta_G(X) = \int_T f(z) \d \mu_G (z) \, . 
\end{equation}
For reference, the measures $\d \mu_G (z)$ for orthogonal and symplectic groups, along with our conventions for Lie algebras, are gathered in Tables \ref{simpleRootsTable} and \ref{measures}. 

\begin{table}[t]
    \centering
    \begin{tabular}{|c|c|}
    \hline 
    $G$ & Simple Roots  \\  \hline  
        $B_r$ & 
        \begin{tabular}{c}
             $\alpha_l = \varepsilon_l - \varepsilon_{l+1}$ ($l<r$)  \\
            $\alpha_r = \varepsilon_r$  
        \end{tabular}
         \\ \hline 
        $C_r$ & 
        \begin{tabular}{c}
             $\alpha_l = \varepsilon_l - \varepsilon_{l+1}$ ($l<r$)  \\
            $\alpha_r = 2 \varepsilon_r$  
        \end{tabular}
         \\ \hline 
        $D_r$ & 
        \begin{tabular}{c}
             $\alpha_l = \varepsilon_l - \varepsilon_{l+1}$ ($l<r$)  \\
            $\alpha_r = \varepsilon_{r-1} + \varepsilon_r$  
        \end{tabular}
         \\
         \hline 
    \end{tabular}
    \caption{Simple Roots for Lie algebras of type $B$, $C$ and $D$ of rank $r$ expressed in an orthonormal basis $(\varepsilon_1 , \dots , \varepsilon_r)$ of $\mathbb{R}^r$. }
    \label{simpleRootsTable}
\end{table}

\begin{table}[t]
    \centering
    \begin{tabular}{|c|c|c|}
    \hline 
    $G$ & Positive roots & $\d \mu_{G} (z)$ \\  \hline  
        $B_r$ & \begin{tabular}{c}
             $\varepsilon_k \pm \varepsilon_l$ ($1 \leq k<l \leq r$)  \\
            $\varepsilon_k$ ($1 \leq k  \leq r$)  
        \end{tabular} & $\left( \prod\limits_{l=1}^r  \frac{\d z_l}{2 \pi i z_l} \right) \prod\limits_{1 \leq k < l \leq r} (1-z_k z_l) \left( 1- \frac{z_k}{z_l}\right) \prod\limits_{l=1}^r \left( 1- z_l\right) $\\ \hline 
        $C_r$ & \begin{tabular}{c}
             $\varepsilon_k \pm \varepsilon_l$ ($1 \leq k<l \leq r$)  \\
            $2 \varepsilon_k$ ($1 \leq k  \leq r$)  
        \end{tabular} &  $\left( \prod\limits_{l=1}^r  \frac{\d z_l}{2 \pi i z_l} \right) \prod\limits_{1 \leq k < l \leq r} (1-z_k z_l) \left( 1- \frac{z_k}{z_l}\right) \prod\limits_{l=1}^r \left( 1- z_l^2\right) $ \\ \hline 
        $D_r$ & \begin{tabular}{c}
             $\varepsilon_k \pm \varepsilon_l$ ($1 \leq k<l \leq r$) 
        \end{tabular} & $\left( \prod\limits_{l=1}^r  \frac{\d z_l}{2 \pi i z_l} \right) \prod\limits_{1 \leq k < l \leq r} (1-z_k z_l) \left( 1- \frac{z_k}{z_l}\right)$ \\
         \hline 
    \end{tabular}
    \caption{Measures on maximal tori of various algebras appearing in the Weyl integration formula.}
    \label{measures}
\end{table}

\subsection{The Weyl formula for non-connected groups}

We now review how the Weyl formula (\ref{WeylFormula2}) is modified when non-connected groups are considered. We follow the construction of \cite{wendt2001weyl}, assuming for simplicity that the connected semisimple Lie group $G$ is of type $A$ with odd rank\footnote{For groups of type $A$ with even ranks there are additional subtleties, see \cite{wendt2001weyl}. }, $D$ or $E$. Let $\Gamma$ be the group of automorphisms of the Dynkin diagram of $G$, so that the group of automorphisms of the root system is $W \rtimes \Gamma$. It is well-known (see Theorem 7.8 in \cite{knapp2013lie}) that $\Gamma$ is also the group of outer automorphisms of the Lie algebra $\mathfrak{g}$ of $G$, and this can be used to construct a homomorphism $\varphi : \Gamma \rightarrow \mathrm{Aut}(G)$. Finally, $\varphi$ allows to define the \emph{principal extension}\footnote{See section I.3 in \cite{Siebenthal} for more about principal extensions. } $\tilde{G} = G \rtimes_{\varphi} \Gamma$ of $G$. This extension contains $|\Gamma|$ connected components. When $\Gamma$ is non-trivial, we then have constructed a non-connected group $\tilde{G}$, and we will now explain, following \cite{wendt2001weyl}, how to integrate a conjugation invariant function over $\tilde{G}$. In subsection \ref{subsection:O(2n)}, we will illustrate this mathematical construction in the case of $G=\mathrm{SO}(2n)$, whose principal extension will appear to be the full $\mathrm{O}(2n)$ group. 

For $\tau \in \Gamma$, we will write the Weyl integration formula for the connected component $G \tau$. We first define the subgroup $S_0(\tau)$ of the maximal torus $T$ that consists of elements left invariant under the action of $\tau$. This plays the role of the maximal torus for the component $G \tau$; more precisely, every element of $G \tau$ is conjugate under $G$ to an element of $S_0(\tau) \tau$. We can then construct the analogous of the map (\ref{psimap}), 
\begin{eqnarray}
     \psi_{\tau} &:& G/S_0(\tau) \times S_0(\tau) \rightarrow G\tau \\
     & & (y,z) \mapsto y z \tau y^{-1} \, .  \nonumber
\end{eqnarray}
As is clear from its definition, $S_0(\tau)$ has in general dimension smaller than the dimension of $T$. We compute the Jacobian of the change of variables as in (\ref{detDiff1}),
\begin{equation}
     \det (\d \psi_{\tau})_{(y,z)} = \det \left( \mathrm{Ad}(\tau^{-1}z^{-1})-1 \right) |_{\mathfrak{s}_0(\tau)^{\perp}} \, , 
\end{equation}
where $\mathfrak{s}_0(\tau)$ is the Lie algebra of $S_0(\tau)$ and $\mathfrak{s}_0(\tau)^{\perp}$ denotes its complement. In this determinant, we have three kinds of contributions: 
\begin{itemize}
    \item If a root $\alpha$ is fixed by $\tau$, i.e. $\tau (\alpha) = \alpha$, then it contributes a factor $(z^{-\alpha} -1)$. 
    \item If a root $\alpha$ is not fixed by $\tau$, i. e. $\tau (\alpha) \neq \alpha$, then the matrix of $\mathrm{Ad}(\tau^{-1}z^{-1})-1$ in the space $\mathfrak{g}_{\alpha} \oplus \mathfrak{g}_{\tau(\alpha)}$ is 
    \begin{equation}
    \label{matrixNotFixedRoots}
        \begin{pmatrix}
        -1 & z^{-\tau(\alpha)} \\ z^{-\alpha} & -1 
        \end{pmatrix} \, , 
    \end{equation}
    and therefore the pair $(\alpha , \tau (\alpha))$ contributes $1-z^{-2\beta}$, where $\beta = \frac{1}{2} (\alpha + \tau (\alpha))$. 
    \item One can show \cite{wendt2001weyl} that the part $\mathfrak{t} \cap \mathfrak{s}_0(\tau)^{\perp}$ of the Cartan only gives a constant, that will be absorbed in the normalization of (\ref{WendtFormula}). 
\end{itemize}
In other words, 
\begin{equation}
     \det \left( \mathrm{Ad}(z^{-1})-1 \right) |_{\mathfrak{s}_0(\tau)^{\perp}} = \prod\limits_{\alpha \in \Delta (\tau)^{\vee}} (1-z^{-\alpha})
\end{equation}
where $\Delta (\tau)$ is the root system obtained by projecting the root system $\Delta$ on the subspace invariant under $\tau$ (this can also be seen as restricting the roots initially defined on $\mathfrak{t}$ to $\mathfrak{s}_0(\tau)$), and ${}^{\vee}$ denotes the Langlands dual (which amounts to doubling the length of the projection of the non-invariant roots). One can prove that this root system is the one obtained after folding the Dynkin diagram of $G$ and taking the Langlands dual. We finally obtain the generalized Weyl formula \cite{wendt2001weyl}
\begin{equation}
     \int_{G\tau} \d \eta_G(X) f(X) = \frac{1}{|W(\tau)|} \int_{S_0 (\tau)} \d \eta_{S_0 (\tau)} (z) f(z \tau ) \prod\limits_{\alpha \in \Delta (\tau)^{\vee}} (1-z^{-\alpha}) \, . 
\end{equation}
and using a reasoning similar to the one presented in Appendix \ref{appendixMeasures} we obtain 
\begin{equation}
\label{WendtFormula}
     \int_{G\tau} \d \eta_G(X) f(X) = \int_{S_0 (\tau)} \d \eta_{S_0 (\tau)} (z) f(z \tau ) \prod\limits_{\alpha \in \Delta (\tau)^{\vee}_+} (1-z^{-\alpha}) \, . 
\end{equation}

\subsection{\texorpdfstring{The $\mathrm{O}(2n)$ integration formula}{}}
\label{subsection:O(2n)}

We now explain how the material of the previous sections will allow us to integrate over the orthogonal group $\mathrm{O}(2n)$.\footnote{The case of $\mathrm{O}(2n+1)$ is much easier; see (\ref{OnoddIntegration}). } If we start with $G=\mathrm{SO}(2n)$, then $\Gamma = \{ 1 , \mathcal{P} \}$, where $\mathcal{P}$ acts on the $D_n$ Dynkin diagram as:\footnote{For $n=4$, $\Gamma$ is in fact the permutation group $\mathfrak{S}_3$, but we will still consider the subgroup of $\mathfrak{S}_3$ generated by any permutation of two of the three external roots, and call it $\Gamma$. } 
    \begin{equation}
    \label{DynkinDiag}
        \begin{tikzpicture}
\node[smallnode] (1) at (0,0) {};
\node[smallnode] (2) at (1,0) {};
\node (3) at (2,0) {$\cdots$};
\node[smallnode] (4) at (3,0) {};
\node[smallnode] (5) at (4,1) {};
\node[smallnode] (6) at (4,-1) {};
\draw[-] (1) -- (2);
\draw[-] (2) -- (3);
\draw[-] (3) -- (4);
\draw[-] (4) -- (5);
\draw[-] (4) -- (6);
\draw[<->] (4.4,.8) to [bend left] node[right] {$\mathcal{P}$} (4.4,-.8);
\end{tikzpicture} 
    \end{equation}
One can check that the action of $\mathcal{P}$ on the group $\mathrm{SO}(2n)$ in its fundamental representation can be realized by the conjugation by the block-diagonal matrix 
\begin{equation}
\label{defP}
     P = \mathrm{Diag} \left( \begin{pmatrix} 
    1 & 0 \\ 0 & 1 
    \end{pmatrix} , \dots , \begin{pmatrix} 
    1 & 0 \\ 0 & 1 
    \end{pmatrix} , \begin{pmatrix} 
    0 & 1 \\ 1 & 0 
    \end{pmatrix}\right) \, . 
\end{equation}
We can now check that the principal extension of $\mathrm{SO}(2n)$ by $\Gamma$ is precisely $\mathrm{O}(2n)$: 
\begin{equation}
\label{isomorphismSOGammaO}
   \widetilde{SO}(2n) = \mathrm{SO}(2n) \rtimes_{\varphi} \Gamma \cong \mathrm{O}(2n) \, . 
\end{equation}
For that, consider the map 
\begin{eqnarray*}
     \Psi &:& \mathrm{SO}(2n) \rtimes_{\varphi} \{ 1 , P \} \rightarrow \mathrm{O}(2n)  \\
     & & (X,\epsilon) \mapsto X \epsilon \, ,  
\end{eqnarray*}
and check that this is indeed a group homomorphism,\footnote{$\Psi ((X,\epsilon)\cdot (X' , \epsilon ')) = \Psi ((X \epsilon X' \epsilon , \epsilon \epsilon ')) = X \epsilon X' \epsilon  \epsilon \epsilon ' = X \epsilon X' \epsilon ' = \Psi ((X,\epsilon)) \Psi ((X',\epsilon')) $. } which is injective and surjective. Therefore, we have made it clear that the construction of the previous subsection which aims at obtaining a Weyl integration formula on the principal extension $\tilde{G}$ reduces for $G=\mathrm{SO}(2n)$ to finding an integration formula for $\tilde{G}=\mathrm{O}(2n)$. 

\begin{figure}
    \centering
    \includegraphics[width=.72\paperwidth]{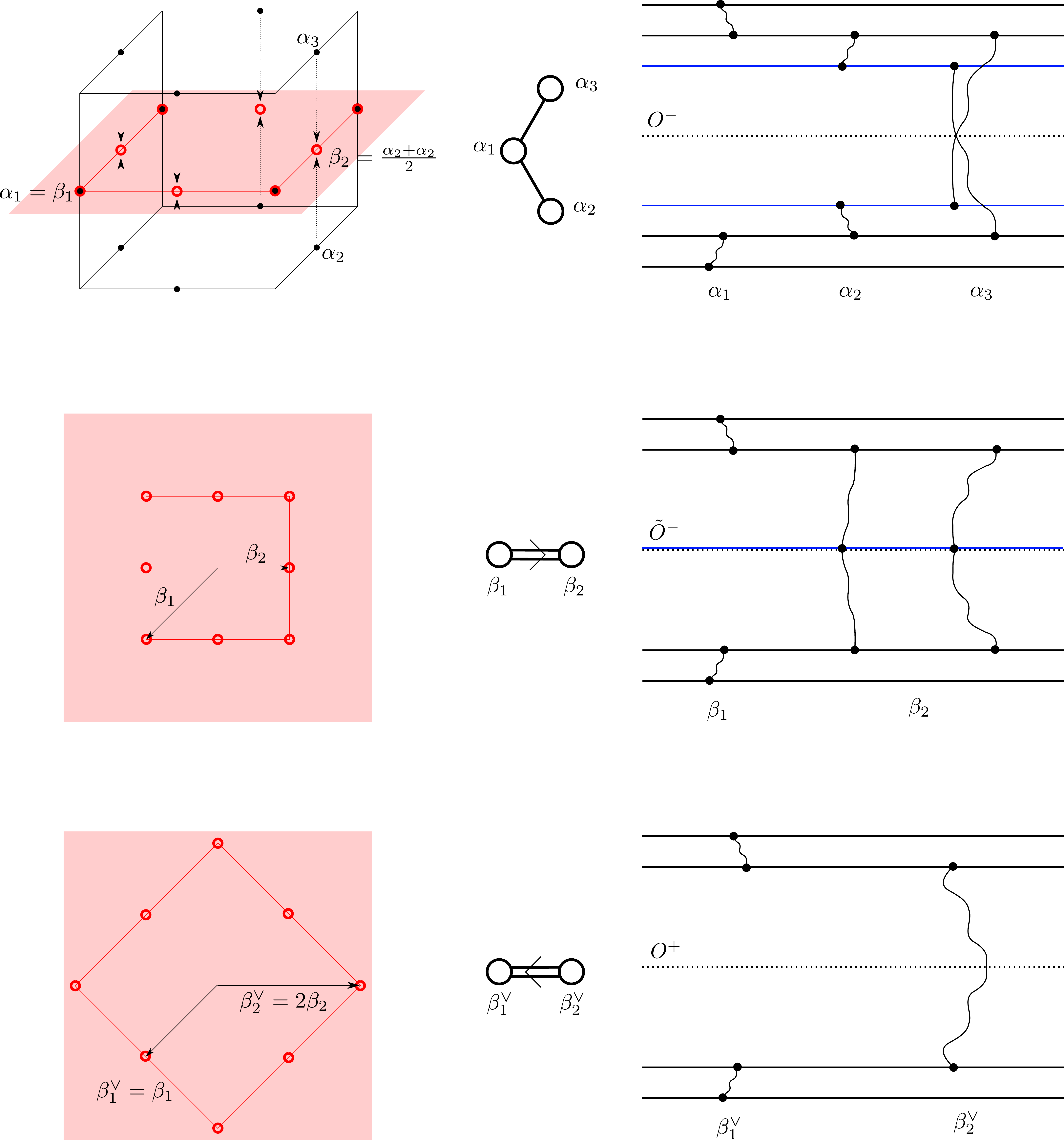}
    \caption{In the upper-left image, the black dots represent the roots of $D_3$, and $\alpha_{1}$, $\alpha_{2}$ and $\alpha_{3}$ are the simple roots. The red plane is the plane fixed by the involution $\mathcal{P}$ which exchanges $\alpha_2$ and $\alpha_3$. The red circles represent the projected roots. They form a $B_2$ system, with simple roots $\beta_{1}$ and $\beta_{2}$, represented below. Finally, the doubling of the short roots due to (\ref{matrixNotFixedRoots}) leads to the lower part of the figure with the $B_2^{\vee}=C_2$ root system. On the right are represented the same steps with branes (full lines) and orientifolds (dotted lines). For the conventions used in the names of the orientifolds, see \cite{Hanany:2000fq}. The black dots represent fundamental strings (wiggly lines) ending on branes. Only strings corresponding to simple roots are represented. The blue branes are those permuted by $\mathcal{P}$. When restricting to the $\mathcal{P}$-invariant configuration, they are replaced by a single brane stuck on the orientifold. In the last part, the $\tilde{O}^-$ thus obtained is replaced by an $O^+$ on which strings can not end \cite{Hanany:2001iy}.  }
    \label{figD3}
\end{figure}

Let us now illustrate this formula in the case $G=\mathrm{SO}(2n)$, using the isomorphism (\ref{isomorphismSOGammaO}). The simple roots can be written in an orthonormal basis $(\varepsilon_1 , \dots , \varepsilon_n)$ as indicated in Table \ref{simpleRootsTable}. The outer automorphism $P$ exchanges $\alpha_{n-1}$ and $\alpha_n$, so the invariant space is the hyperplane spanned by $(\varepsilon_1 , \dots , \varepsilon_{n-1})$. One concludes that the roots $\alpha_i$ for $1 \leq i \leq n-2$ are left invariant, while $\alpha_{n-1}$ and $\alpha_n$ combine to give, after doubling the length, a long root. Summarizing, we have $\Delta (1)^{\vee} = D_n$ and $\Delta (\mathcal{P})^{\vee} = C_{n-1}$.\footnote{Note that some connections between characters of $\mathrm{O}_-(2n)$ and $\mathrm{Sp}(n-1)$ have been noticed previously \cite{Hanany:2016gbz}. } This is illustrated in the particular case of $D_3$ in the left part of Figure \ref{figD3}, in which we also include on the right side a string theory picture with branes and orientifold planes of the same process. We conclude that the final formula for integrating a conjugation-invariant function over $\mathrm{O}(2n)$ is 
\begin{equation}
\label{OnIntegration}
     \int_{\mathrm{O}(2n)} \d \eta_{\mathrm{O}(2n)}(X) f(X) = \frac{1}{2} \left[  \int \d \mu_{\mathrm{SO}(2n)} (z) f(z) + \int  \d \mu_{\mathrm{Sp}(n-1)} (z) f(z \mathcal{P} )  \right]  \, . 
\end{equation}
In section \ref{subsectionHiggsBranchComputation}, we will make use of these formulas in a Hilbert series computation, where we will explain how to concretely take care of the operator $\mathcal{P}$. For completeness, we also give the analogous formula for an $\mathrm{O}(2n+1)$ integration: 
\begin{equation}
\label{OnoddIntegration}
     \int_{\mathrm{O}(2n+1)} \d \eta_{\mathrm{O}(2n+1)}(X) f(X) = \frac{1}{2} \left[  \int \d \mu_{\mathrm{SO}(2n+1)} (z) f(z) + \int  \d \mu_{\mathrm{SO}(2n+1)} (z) f(-z)  \right]  \, . 
\end{equation}

\section{\texorpdfstring{Moduli space of $k$ $\mathrm{Sp}(N)$ instantons}{}}
\label{sectionModuliSpace}

 In this section we apply the formula (\ref{OnIntegration}) in the context of the moduli space of $k$ $\mathrm{Sp}(N)$ instantons. In general the moduli space of instantons on $\mathbb{R}^4$ can be studied using the ADHM constriction \cite{adhm}. At later stage it was understood that such a mathematical construction can be embedded in a $4d$ $\mathcal{N}=2$ quiver gauge theory  \cite{Douglas:1996uz,Witten:1995gx,Witten:1994tz,Douglas:1995bn}. From a QFT point of view the the instanton moduli space is realized as the Higgs branch of the quiver gauge theory taken in consideration.  Moreover since the Higgs branch receives no quantum corrections \cite{Argyres:1996eh} the moduli space of $k$ instantons can be realized as the Higgs branch of the corresponding three-dimensional $\mathcal{N}=4$ theory. For the particular case of $k$ $\mathrm{Sp}(N)$ instantons the quiver gauge that describes the moduli space of $k$ $\mathrm{Sp}(N)$ instantons is realized by the following quiver diagram \cite{Cremonesi:2014xha} \footnote{Note that here we are using $\mathcal{N}=4$ notation. }
\begin{equation}
        \label{HiggsQuiver}
        \begin{tikzpicture}
\node[roundnode] (1) {$\mathrm{O}(k)$};
\node[squarednode] (2) [right=of 1] {$\mathrm{Sp}(N)$};
\draw (1) edge[loop left] node {Sym} (1);
\draw[-] (1) -- (2);
\end{tikzpicture}  
    \end{equation}
Alternatively using mirror symmetry \cite{Intriligator:1996ex} the moduli space of $k$ $\mathrm{Sp}(N)$ instantons can be studied using the Coulomb branch of the associated mirror theory. This analysis has been performed systematically in \cite{Cremonesi:2014xha}, which we use in section \ref{appendix:coulombbranch} to provide a sharp confirmation of our results. 

\subsection{Higgs branch computation}
\label{subsectionHiggsBranchComputation}

The Higgs branch unrefined Hilbert series of the theory (\ref{HiggsQuiver}) is \cite{derksen2015computational,Hanany:2012dm}
\begin{equation}
\label{HilbertSeriesGeneralExpr}
    g_{N,k}(t) = \int_{\mathrm{O}(k)} \d \eta_{\mathrm{O}(k)} (X)  \frac{\det \left(1-t^2 \Phi_{\mathrm{Adj}} (X) \right)}{\det \left(1-t \Phi_{\mathrm{Fund}} (X) \right)^{2N} \det \left(1-t \Phi_{\mathrm{Sym}} (X) \right)^2} \, . 
\end{equation} 
Here, $\Phi_{\mathrm{Adj}} (X)$, $\Phi_{\mathrm{Fund}} (X)$, and $\Phi_{\mathrm{Sym}} (X)$ denote the image of a generic element $X \in \mathrm{O}(k)$ in the adjoint, fundamental and symmetric representations of $\mathrm{O}(k)$ respectively. 
%The numerator of (\ref{HilbertSeriesGeneralExpr}) accounts for the F-terms, which transform in the adjoint of $\mathrm{O}(k)$, while the denominator is related to the generators of the Higgs branch. 
%Finally I preferedd to eliminate the full sentence. In general it's not true that denominator is related with the generators of the space. This is due the fact that the generators of the space are gauge invariants operators, while the terms at denominator are given by letter counting of the possible operators that can be written starting from the quiver (and do not necessarly correspond to the generators). Well this is true for the connected component (SO(2n)), for the non-connected component, the procedure that we have to use the in order to get these operators is more abstract since we have to play with the matrix P. So I would skip completely this sentence. 
%HOWEVER IF YOU WANT TOMORROW WE CAN REPLACE IT WITH AN OTHER SENTENCE, LET ME KNOW.
Using (\ref{OnIntegration}), we can split this integral in two parts, called $ g_{N,k}^+(t) $ and $ g_{N,k}^-(t)$, given for even $k=2n$ by 
\begin{align}
\label{eq:ghs}
     g_{N,2n}(t) &= \frac{1}{2} \left[ g_{N,2n}^+(t) + g_{N,2n}^-(t) \right] \, ,\\  
     g_{N,2n}^+(t) &= \int \d \mu_{\mathrm{SO}(2n)} (z) \frac{\det \left(1-t^2 \Phi_{\mathrm{Adj}} (z) \right)}{\det \left(1-t \Phi_{\mathrm{Fund}} (z) \right)^{2N} \det \left(1-t \Phi_{\mathrm{Sym}} (z) \right)^2} \, , \\
     g_{N,2n}^-(t) &= \int  \d \mu_{\mathrm{Sp}(n-1)} (z) \frac{\det \left(1-t^2 \Phi_{\mathrm{Adj}} (z) \Phi_{\mathrm{Adj}} (\mathcal{P})  \right)}{\det \left(1-t \Phi_{\mathrm{Fund}} (z)   \Phi_{\mathrm{Fund}} (\mathcal{P})  \right)^{2N} \det \left(1-t \Phi_{\mathrm{Sym}} (z)  \Phi_{\mathrm{Sym}} (\mathcal{P})  \right)^2}  \, , 
\end{align}
where the fugacities $z_i$ parameterize the fundamental tori while the $t$ fugacity accounts for the conformal dimension.
The measures $\d \mu_{\mathrm{SO}(2n)} (z)$ and $\d \mu_{\mathrm{Sp}(n-1)} (z)$ can be read directly from Table \ref{measures}. To be concrete, we choose for $\Phi_{\mathrm{Fund}} (z)$ the diagonal matrix 
\begin{equation}
\label{matrixz}
     \Phi_{\mathrm{Fund}} (z) = \mathrm{Diag} \left( z_1 , \frac{1}{z_1} , \dots , z_n , \frac{1}{z_n} \right) \, , 
\end{equation}
and $\Phi_{\mathrm{Fund}} (\mathcal{P}) = P$, where the matrix $P$ was defined in equation (\ref{defP}). Once these arbitrary choices have been made, we can deduce the matrix form of $z$ and $\mathcal{P}$ in the adjoint (or antisymmetric) and symmetric representations. The details about how this is done, along with an explicit example in the cases $2n=2$ and $2n=4$, can be found in appendix \ref{subsectionRepresentations}.

\subsection{Case studies}
In this subsection we report the results obtained for $k=2$ and $k=4$, both with $N=1$. We collect further results for higher values of $k$ and $N$ in Appendix \ref{appendix:hilbertseries}.

\subsubsection{\texorpdfstring{Two  $\mathrm{Sp}(1)$ instantons}{}}

Let's review with full details the $\mathrm{O}(2)$ case with two flavours. This case was already considered in \cite{Hanany:2012dm}. The formula (\ref{eq:ghs}) leads to\footnote{The Plethystic Exponential (PE) of a function $f(\textbf{x})$ such that $f(0)=0$ is defined as \cite{Benvenuti:2006qr}
\begin{equation}
\textrm{PE}[f(\textbf{x})] = \textrm{Exp}\left[\sum_{n=1}^{\infty}\frac{f(\textbf{x}^n)}{n}\right]\, ,    
\end{equation}
where $\textbf{x}$ denotes the full set of all fugacities on which the function $f$ depends.
} 
\begin{equation}\begin{split}
g_{1,2}^+(t) & = \oint_{\mid z_1\mid=1} \frac{dz_1}{z_1}(1-t^2)\textrm{PE}\left[2\left(z_1^2+1+\frac{1}{z_1^2}\right)t+2\left(z_1+\frac{1}{z_1}\right)t\right] =\\
& = \frac{1-t+5 t^2+4 t^3+4 t^4+4 t^5+5 t^6-t^7+t^8}{(1-t)^8 (1+t)^2 \left(1+t+t^2\right)^3} \, ,
\end{split}\end{equation}
while for the minus component we get
\begin{equation}
g_{1,2}^{-}(t)=\frac{1+t^2}{(1-t)^6(1+t)^4} \, .  
\end{equation}
As it has been previously observed in \cite{Hanany:2012dm} the expression for $g_{1,2}^{-}(t)$ is obtained without performing any integration. Here this fact can be fully understood since emerges as a natural consequence of the application of the formula (\ref{OnIntegration}). Therefore the unrefined Hilbert series reads    
\begin{eqnarray}
    g_{1,2}(t) &=& \frac{1+t+3 t^2+6 t^3+8 t^4+6 t^5+8 t^6+6 t^7+3
   t^8+t^9+t^{10}}{(1-t)^8 (1+t)^4 \left(1+t+t^2\right)^3} \, .  
\end{eqnarray}
This result agrees with the Hilbert series previously found in \cite{Hanany:2012dm}. Finally we perform a power series expansion of the plus and minus components 
\begin{eqnarray}
    g_{1,2}^+(t) &=&  1+2 t+10 t^2+30 t^3+76 t^4+178 t^5 + O\left(t^{5}\right) \, , \nonumber \\
    g_{1,2}(t) &=& 1+2 t+9 t^2+22 t^3+55 t^4+116 t^5 + O\left(t^{5}\right) \, .  
\end{eqnarray}
We observe that the two expansions begin to disagree starting from the order $t^2$. We comment on this point in subsection \ref{subsec:lettercounting} where we give an interpretation in terms of gauge invariant operators.

\subsubsection{\texorpdfstring{Four $\mathrm{Sp}(1)$ instantons}{}} 

The application of the formula (\ref{HilbertSeriesGeneralExpr}) for $n=2$ and $N=1$ gives, using the computations of Appendix \ref{subsectionRepresentations}, 
\begin{eqnarray*}
     g_{1,4}^+(t) &=&  \frac{1}{(1-t)^{16} (1+t)^8 \left(1+t+t^2+t^3+t^4\right)^3
   \left(1+t+2 t^2+t^3+t^4\right)^4} \times \\
   & & (1+t+3 t^{2}+9 t^{3}+22 t^{4}+43 t^{5}+91
   t^{6}+179 t^{7}+355 t^{8}+626 t^{9} \\
   & & +1065 t^{10}+1661
   t^{11}+2471 t^{12}+3425 t^{13}+4504 t^{14}+5525 t^{15}+6425
   t^{16}\\
   & &+6983 t^{17}+7210 t^{18} + \textrm{palindrome} + t^{36}) \, , \\
    g_{1,4}^-(t) &=& \frac{1}{(1-t)^{14} (1+t)^6
   \left(1+t+2 t^2+t^3+t^4\right)^4} \times \\
   & &(1-2 t+5 t^{2}-2 t^{3}+12 t^{4}+2 t^{5}+20
   t^{6}+12 t^{7}+38 t^{8}+14 t^{9}\\
   & &+44 t^{10}+ \textrm{palindrome} + t^{20}) \, , \\
    g_{1,4}(t) &=& \frac{1}{(1-t)^{16} (1+t)^8 \left(1+t^2\right)^4
   \left(1+t+t^2\right)^4 \left(1+t+t^2+t^3+t^4\right)^3} \times\\
   & & (1+t+3 t^2+9 t^{3}+22 t^{4}+43 t^{5}+85
   t^{6}+153 t^{7}+273 t^{8}+440 t^{9}\\
   & &+680 t^{10}+982
   t^{11}+1364 t^{12}+1778 t^{13}+2225 t^{14}+2633 t^{15}+2981
   t^{16}\\
   & &+3187 t^{17}+3274 t^{18} + \textrm{palindrome} + t^{36}) \, .  
\end{eqnarray*}
Let us write down the expansion of the two Hilbert series $g_{1,4}^+(t)$ and $g_{1,4}(t)$ up to the first order which distinguishes them, which happens to be the sixth order:
\begin{eqnarray}
\label{g14expansion}
     g_{1,4}^+(t) &=& 1+2 t+9 t^2+26 t^3+78 t^4+202 t^5+524 t^6+ O\left(t^7\right) \, , \nonumber\\
    g_{1,4}(t) &=& 1+2 t+9 t^2+26 t^3+78 t^4+202 t^5+518 t^6+ O\left(t^7\right) \, . 
\end{eqnarray}
Again, we will explain in the next subsection why the two series agree for small conformal dimension. 

We have computed (\ref{HilbertSeriesGeneralExpr}) for a lot of values of $N$ and $k$, some of which are reported in Appendix \ref{appendix:hilbertseries}. In all cases we observe that the numerator is a palindromic polynomial. This implies that the algebaric variety describing the moduli space of instantons is a Calabi-Yau space \cite{Gray:2008yu,Forcella:2008eh}.
Moreover the dimension of the pole at $t=1$ matches the complex dimension of the corresponding moduli space of instantons. As it is well known the complex dimension of the moduli space of $k$ $G$-instantons is $2kh_G$, where $h_G$ is the dual Coxeter number of $G$. In particular for $G=\mathrm{Sp}(N)$ we have $h_{\mathrm{Sp}(N)}=N+1$.

\subsection{Refined Hilbert Series and Letter Counting}
\label{subsec:lettercounting}

In this subsection, we propose a direct check of our method by comparing the results of the integration (\ref{HilbertSeriesGeneralExpr}) with an explicit counting of gauge invariant operators on the Higgs branch. For this purpose, it is handy to consider the refined Hilbert series to allow for a more precise identification of the operators involved. Thus we introduce the fugacities $x$ and $y_i$ for the $\mathrm{SU}(2)$ and $\mathrm{Sp}(N)$ global symmetries respectively. The Hilbert series (\ref{HilbertSeriesGeneralExpr}) becomes 
\begin{align}
\label{refinedHilbertSeriesIntegration}
 \resizebox{0.97\hsize}{!}{$ g_{N,k}(t;x,\vec{y}) = \int_{\mathrm{O}(k)} \d \eta_{\mathrm{O}(k)} (X)  \frac{\det \left(1-t^2 \Phi_{\mathrm{Adj}} (X) \right)}{\det \left(1-t \Phi_{\mathrm{Fund}} (X) \otimes  \Phi_{\mathrm{Fund}}^{\mathrm{Sp}(N)}(\vec{y}) \right) \det \left(1-t \Phi_{\mathrm{Sym}} (X) \otimes \Phi_{\mathrm{Fund}}^{\mathrm{SU}(2)}(x)\right)}$} \, ,  
\end{align} 
where $\Phi_{\mathrm{Fund}}^{\mathrm{SU}(2)}(x) = \mathrm{Diag} \left( x,x^{-1}\right)$ and similarly $\Phi_{\mathrm{Fund}}^{\mathrm{Sp}(N)}(\vec{y})$ is a diagonal matrix with entries corresponding to the character of the fundamental representation of $\mathrm{Sp}(N)$. We can now use the formula (\ref{OnIntegration}) on (\ref{refinedHilbertSeriesIntegration}). Then in principle we can find a matching with gauge invariant operators built from the fundamentals $Q$ and symmetric tensors $S$ which transform under the various gauge and global symmetry groups as reported in Table \ref{table:fields} \footnote{Henceforth $\alpha,\beta=1,2$ denote $\mathrm{SU}(2)$ indices, $a,b=1, \dots , k$ denote $\mathrm{O}(k)$ indices, and $i,j=1,\dots , 2N$ denote $\mathrm{Sp}(N)$ indices.}. 

\begin{table}[h!]
\centering
    \begin{tabular}{|c|c|c|c|c|}
    \hline
    & $U(1)_t$ & $\mathrm{SU}(2)_x$ & $\mathrm{Sp}(N)_{\vec{y}}$ & $\mathrm{O}(k)_{\vec{z}}$ \\ \hline
     $S^{\alpha}_{ab}$ & 1 & $\mathbf{2}$ & 1 & Symmetric \\
     $Q^i_a$ & 1 & 1 & $\mathbf{2N}$ & $\mathbf{k}$ \\ \hline
    \end{tabular}
    \caption{The charges and representations under which the various fields transform.}
    \label{table:fields}
    \end{table}
In general given a gauge group $G$ the gauge invariant operators $\mathcal{O}(Q,S)$ are those which satisfy
\begin{equation}
    \forall X \in G \, , \qquad \mathcal{O}(Q,S) = \Phi_{\mathcal{O}}(X)\left( \mathcal{O}(Q,S) \right):= \mathcal{O}(XQ,XSX^{-1}) \, . 
\end{equation}
Given an $\mathrm{SO}(k)$ invariant operator $\mathcal{O}(Q,S)$, and any matrix $P \in O_-(k)$, the fact that $P^2 \in \mathrm{SO}(k)$ forces $\Phi_{\mathcal{O}}(P) = \pm 1$. If $\Phi_{\mathcal{O}}(P) = + 1$, then $\mathcal{O}$ is also gauge invariant under $\mathrm{O}(k)$, and in the other case, it is not, and will only be present on the Higgs branch of the $\mathrm{SO}(k)$ theory. This explains why the coefficients of $g_{N,k}(t)$ will always be smaller or equal to the coefficients of $g_{N,k}^+(t)$. 

We now have to enumerate all possible gauge invariant operators constructed out of the fields of Table \ref{table:fields}. This can be reformulated as: find a family of generators of the ring of $\mathrm{O}(k)$ and $\mathrm{SO}(k)$ invariants of the representation 
\begin{equation}
\label{repres}
     \underbrace{\mathrm{Fundamental} \oplus \cdots \oplus \mathrm{Fundamental}}_{2N \textrm{ terms}} \oplus \mathrm{Symmetric} \oplus \mathrm{Symmetric} \, . 
\end{equation}
For a representation made only of copies of the fundamental representation, it is known (see for instance Theorem 11.2.1 in \cite{procesi2006lie}) that the ring of $\mathrm{SO}(k)$ invariants is generated by the contractions $\delta_{ab} Q^a_i Q^b_j$ and by the determinants $\det (Q^a_i) := \epsilon_{a_1 \cdots a_k} Q^{a_1}_{i_1} \cdots  Q^{a_k}_{i_k}$, while the ring of $\mathrm{O}(k)$ invariants is generated by the contractions $\delta_{ab} Q^a_i Q^b_j$ only. Similarly, one can deduce from the results of \cite{aslaksen1995invariant} that for copies of the symmetric representation, the ring of invariants for both $\mathrm{O}(k)$ and $\mathrm{SO}(k)$ is generated by the $\mathrm{Tr}(S^{n})$, for $n \geq 1$. Combining those single-trace generators, one obtains the full spectrum of invariant operators at each level. For our representation (\ref{repres}), we also have to consider single trace operators that mix fundamentals and symmetrics, but the principle is the same: we have to consider all contractions with the tensors $\delta_{ab}$ and $\epsilon_{a_1 \cdots a_k}$. Operators that don't contain the $\epsilon$ tensor are always invariant under $\mathrm{SO}(k)$, although the converse is not true. Hence for a value of the conformal dimension $\leq k$, the series expansions of $g_{0,k}^+(t)$ and $g_{0,k}(t)$ agree, as observed in the previous section. Finally, we have to take into account the relations coming from the vanishing of the F-terms of the quiver theory, and additional algebraic relations. 

A rather trivial consequence of the analysis above is that with no flavor, i.e. $N=0$, the spectrum of operators is the same for $\mathrm{O}(k)$ and $\mathrm{SO}(k)$ gauge groups. So we must have 
\begin{equation}
    g_{0,k}^+(t) = g_{0,k}(t) \, ,  
\end{equation}
for all $k$. This can be checked explicitly with the integration formula. Physically, these are the Hilbert series for the moduli space of $k$ rank zero instantons, and as such they can be obtained from the $g_{0,1}(t)$ function by plethystic exponentiation: 
\begin{equation}
    \mathrm{PE} \left[ g_{0,1}(t) \nu \right] = \sum\limits_{k=0}^{\infty} g_{0,k}(t)\nu^k \, . 
\end{equation}
One can readily evaluate\footnote{Using $  \Phi_{\mathrm{Sym}} (1) = \Phi_{\mathrm{Sym}} (\mathcal{P}) = 1$ and $  \Phi_{\mathrm{Fund}} (1) = - \Phi_{\mathrm{Fund}} (\mathcal{P}) = 1$ in (\ref{OnoddIntegration}). }  \cite{Benvenuti:2010pq}
\begin{equation}
   g_{N,1}(t) =  \frac{1}{2(1-t)^2} \left( \frac{1}{(1-t)^{2N}} + \frac{1}{(1+t)^{2N}} \right) \, ,  
\end{equation}
from which, setting $N=0$,  all the $g_{0,k}(t)$ can be obtained. 

\begin{table}
    \centering
    \begin{tabular}{|c|c|c|}
    \hline
        Order & $\mathcal{O}$ such that $\Phi_{\mathcal{O}}(P) = +1$ & $\mathcal{O}$ such that $\Phi_{\mathcal{O}}(P) = -1$ \\ \hline
        $t^0$ & $ 1 \rightarrow [0,0] $ & $-$ \\ \hline
        $t^1$ & $\mathrm{Tr}(S^{\alpha})\rightarrow  [1,0] $ & $-$ \\ \hline
        $t^2$ & 
        \begin{tabular}{c}
             $(Q^T)^i Q^j \rightarrow [0,2]$  \\
             $\mathrm{Tr}(S^{\alpha}) \mathrm{Tr}(S^{\beta}) \rightarrow [2,0]$ \\
             $\mathrm{Tr}(S^{\alpha}S^{\beta}) \rightarrow [2,0]$
        \end{tabular}
         & $\mathrm{det} (Q) \rightarrow [0,0]$ \\ \hline
         $t^3$ & 
        \begin{tabular}{c}
             $\mathrm{Tr}(S^{\alpha}S^{\beta} S^{\gamma}) \rightarrow [3,0]$ \\
             $\mathrm{Tr}(S^{\alpha}S^{\beta}) \mathrm{Tr}(S^{\gamma}) \rightarrow [3,0] + [1,0]$ \\
             $(Q^T)^i S^{\alpha} Q^j \rightarrow [1,2]$ \\
             $(Q^T)^i Q^j \mathrm{Tr}(S^{\alpha}) \rightarrow [1,2]$ 
        \end{tabular}
         &   \begin{tabular}{c}
             $(Q^T)^{(i} \epsilon S^{\alpha} Q^{j)} \rightarrow [1,2]$ \\
             $\mathrm{det} (Q) \mathrm{Tr}(S^{\alpha}) \rightarrow [1,0] $ 
        \end{tabular}
         \\ \hline
    \end{tabular}
    \caption{List of operators contributing to the Higgs branch Hilbert series for the $\mathrm{SO}(2)$ theory (columns 2 and 3) and for the $\mathrm{O}(2)$ theory (column 2 only). Our notation is that $\mathrm{Tr}$ means a contraction with a Kronecker $\delta_{ab}$ acting on the gauge indices, while the determinant stands for a contraction with the completely antisymmetric tensor $\epsilon_{ab}$. For instance $\mathrm{det} (Q) = \epsilon_{ab} Q^i_a Q^j_b$. The parenthesis used on indices stand for the symmetrization, for instance $(Q^T)^{(i} \epsilon S^{\alpha} Q^{j)} = \frac{1}{2} \left( (Q^T)^{i} \epsilon S^{\alpha} Q^{j} + (Q^T)^{j} \epsilon S^{\alpha} Q^{i} \right)$. Finally, note that some operators are absent because of algebraic relations, for instance $(Q^T)^i \epsilon S^{\alpha} \epsilon Q^j$ is not included because it is equal to $(Q^T)^i S^{\alpha} Q^j - (Q^T)^i Q^j \mathrm{Tr}(S^{\alpha})$.  }
    \label{tableOperatorsSO2}
\end{table}

As a less trivial illustration, let us consider the example of $N=1$ and $k=2$. The exact refined Hilbert series can be easily obtained exactly. However their expressions are not very illuminating, so we only give the beginning of their power series expansion written in terms of $\mathrm{SU}(2) \times \mathrm{Sp}(1)$ characters, 
\begin{eqnarray*}
     g_{1,2}^+(t) &=& 1+ \chi _{[1,0]}t+ \left(\chi _{[0,0]}+\chi _{[0,2]}+2 \chi _{[2,0]}\right)t^2+ \left(2 \chi _{[1,0]}+3 \chi _{[1,2]}+2 \chi _{[3,0]}\right)t^3+O\left(t^4\right) \\
    g_{1,2}(t) &=& 1+ \chi _{[1,0]}t+ \left(\chi _{[0,2]}+2 \chi _{[2,0]}\right)t^2+ \left(\chi _{[1,0]}+2 \chi _{[1,2]}+2 \chi _{[3,0]}\right)t^3+O\left(t^4\right) \, ,   
\end{eqnarray*}
where 
\begin{equation}
    \chi_{[n,m]} = (x^{-n}+ x^{-n+2} + \dots + x^n)(y^{-m}+ y^{-m+2} + \dots + y^m) \, . 
\end{equation}
The corresponding operators are listed in Table \ref{tableOperatorsSO2}, where our notations are explicited. One finds a perfect match in both cases. 

\subsection{Coulomb branch computation}
\label{appendix:coulombbranch}

In \cite{Cremonesi:2014xha} it has been proved using mirror symmetry that the moduli space of  $k$ $\mathrm{Sp}(N)$ instantons on $\mathbb{C}^2$ can be realized either as the Higgs branch of the 3d $\mathcal{N}=4$ $\mathrm{O}(k)$ gauge theory with one symmetric tensor and a flavor group $\mathrm{Sp}(N)$, represented by the quiver (\ref{HiggsQuiver}) or as the Coulomb branch of the following 3d $\mathcal{N}=4$ quiver gauge theory (with $N+1$ $U(k)$ nodes):
    \begin{equation}
    \label{CoulombQuiverGeneral}
        \begin{tikzpicture}
\node[roundnode] (1) {$U(1)$};
\node[roundnode] (2) [right=of 1] {$U(k)$};
\node[roundnode] (3) [right=of 2] {$U(k)$};
\node (4) [right=of 3] {$\cdots$};
\node[roundnode] (5) [right=of 4] {$U(k)$};
\node[roundnode] (6) [right=of 5] {$U(k)$};
\draw[-] (1) -- (2);
\draw[->,double distance=2pt] (2) -- (3);
\draw[-] (3) -- (4);
\draw[-] (4) -- (5);
\draw[<-,double distance=2pt] (5) -- (6);
\end{tikzpicture} \, . 
    \end{equation}
For the particular case $N=1$ it is natural to consider the following quiver \footnote{That is due to the fact that $C_1$ and $A_1$ are the same algebras.}  
 \begin{equation}
 \label{quiverNequals1}
        \begin{tikzpicture}
\node[roundnode] (1) {$U(1)$};
\node[roundnode] (2) [right=of 1] {$U(k)$};
\node[roundnode] (3) [right=of 2] {$U(k)$};
\draw[-] (1) -- (2);
\draw[-] (2) to [bend right] (3);
\draw[-] (2) to [bend left] (3);
\end{tikzpicture} \,  
    \end{equation}
which is the over-extended Dynkin diagram for $A_1$. 

Using the monopole formula introduced in \cite{Cremonesi:2013lqa}, we can readily write the formula for the Coulomb branch Hilbert series of the theories represented by the quivers (\ref{quiverNequals1}) for $N=1$ and by the quiver (\ref{CoulombQuiverGeneral}) for $N \geq 2$. We label each gauge node with an index $\alpha=0,...,N$ and to each of them we associate a diagonal magnetic flux $\vec{m}^\alpha = \left(m^\alpha_i \right)_{i=1 , \dots , k}$. The monopole formula for the Couloumb branch Hilbert series $HS_{N,k}$ for the case at hand reads \cite{Cremonesi:2014xha}
\begin{equation}
\label{CoulombBranchSeries}
     HS_{N,k}(t) = \sum t^{2\Delta_{N,k} (\vec{m}^0 , \dots , \vec{m}^{N})} \prod\limits_{\alpha=0}^{N} P_U (t^2 ,\vec{m}^{\alpha}) \, , 
\end{equation}
where the sum runs over all fluxes $(m_i^{\alpha})$ such that $m_1^{\alpha} \geq \dots \geq m_k^{\alpha}$ for all $\alpha$, and where the function $\Delta_{N,k}$ reads
\begin{equation}
    \Delta_{N,k} (\vec{m}^0 , \dots , \vec{m}^{N}) = \frac{1}{2}\sum\limits_{i=1}^k |m^0_i|+ \frac{1}{2} \sum\limits_{i,j=1}^k \sum\limits_{\alpha=0}^{N-1} |\ell(\alpha) m^{\alpha}_i - \ell(\alpha+1) m^{\alpha+1}_j| -  \sum\limits_{1 \leq j<i \leq k} \sum\limits_{\alpha=0}^{N} |m^{\alpha}_i - m^{\alpha}_j|\, ,
\end{equation}
where $\ell(\alpha)=1$ for $0<\alpha<N$ , while $\ell (0) = \ell(N)=2$. Finally the factors $P_U (t^2 ,\vec{m}^{\alpha} ) $ are given by \cite{Cremonesi:2013lqa}
\begin{equation}
    P_U (t^2 ,\vec{m} ) = \prod\limits_{j=1}^k \prod\limits_{i=1}^{\lambda_j (\vec{m})} \frac{1}{1-t^{2i}} \, ,
\end{equation}
where $\sum \lambda_j (\vec{m}) = k$ is a partition of $k$ which encodes how many of the various $m_i$ are equal. 

Therefore using the formula (\ref{CoulombBranchSeries}) and mirror symmetry we can check the results obtained in section \ref{subsectionHiggsBranchComputation}. The Coulomb branch computation, even for low values of the gauge groups rank $k$, is quite involved. Nevertheless, although we were not able to sum up the series (\ref{CoulombBranchSeries}) for $ k \geq 4$, we can expand it as a series in $t$ using a brute force computation. For instance one obtains 
\begin{equation}
    HS_{1,4}(t) = 1+2 t+9 t^2+26 t^3+78 t^4+202 t^5+518 t^6+ 1228 t^7 + o(t^7) \, ,
\end{equation}
and we find perfect agreement with the $g_{1,4}(t)$ expansion given in (\ref{g14expansion}). We performed similar checks for other values of $k$ and $N$.

\section{Conclusion}

We have reviewed how to integrate class functions on certain non-connected Lie groups and derived an explicit formula for the case of $\mathrm{O}(2n)$. This has then allowed us to compute the Hilbert series for the  moduli spaces of $k$ $\mathrm{Sp}(N)$ instantons on $\mathbb{R}^4$, thus filling a gap in the literature. As a check of our method and our computations, we have evaluated numerically the Coulomb branch Hilbert series of the corresponding mirror theory. 

This work leaves open several interesting questions. As for all Hilbert series for instanton moduli space, our results can always be written as the quotient of a palindromic polynomial by a polynomial with roots on the unit circle. It would be interesting to derive a general formula for those polynomials, since the computational cost grows fast as the number of instantons and the rank of the group are increased. 

In the light of this work, we have an efficient tool to characterize aspects of theories with non-connected gauge groups, which can be applied in many other situations that orthogonal groups. In particular, one can define new theories based on principal extension gauge groups $\tilde{G}$, which are studied in \cite{Bourget:2018ond}.

\subsection*{Acknowledgements}

It is a pleasure to thank Diego Rodríguez-Gómez for helpful comments and discussions. We would like to acknowledge support from the grant ANR-13-BS05-0001, the EU CIG grant UE-14-GT5LD2013-618459 as well as Asturias Government grant FC-15-GRUPIN14-108 and Spanish Government grant MINECO-16-FPA2015-63667-P. Moreover A.P. is supported by the Asturias Government SEVERO OCHOA grant BP14-003.

\appendix

\section{Some Representations of Orthogonal Groups}
\label{appendix:representations}

\label{subsectionRepresentations}

Although the considerations reviewed here are quite generic, we will focus on representations of orthogonal groups, since this is what we make use of in this article. In the following, $G$ will be an orthogonal group $\mathrm{O}(n)$ or $\mathrm{SO}(n)$ with $n \geq 1$. 

The fundamental representation is the map 
\begin{eqnarray}
      \Phi_{\textrm{Fund}} &:& G \rightarrow \mathrm{Aut}(\mathbb{R}^n) \\
      & & X \mapsto \left[ v \mapsto Xv \right] \, .  \nonumber
\end{eqnarray}
In other words, $\Phi_{\textrm{Fund}} (X)$ acts on $\mathbb{R}^n$ by matrix multiplication on the left. 

The symmetric and antisymmetric (which in the case of orthogonal groups is the adjoint) representations are the maps 
\begin{eqnarray}
      \Phi_{\textrm{Sym}} &:& G \rightarrow \mathrm{Aut}\left( (\mathbb{R}^n \otimes \mathbb{R}^n )/ \mathfrak{S}_2 \right) \\
      & & X \mapsto \Phi_{\textrm{Sym}} (X) \, ,  \nonumber
\end{eqnarray}
and 
\begin{eqnarray}
      \Phi_{\textrm{Adj}} &:& G \rightarrow \mathrm{Aut}\left(\bigwedge\nolimits^2 \mathbb{R}^n \right) \\
      & & X \mapsto \Phi_{\textrm{Adj}} (X) \, , \nonumber
\end{eqnarray}
defined by their actions on a basis $(\varepsilon_k \otimes \varepsilon_l)_{1 \leq k \leq l \leq n}$ of $(\mathbb{R}^n \otimes \mathbb{R}^n )/ \mathfrak{S}_2$ and $(\varepsilon_k \wedge \varepsilon_l)_{1 \leq k < l \leq n}$ of $\bigwedge\nolimits^2 \mathbb{R}^n$ respectively by 
\begin{eqnarray}
        \Phi_{\textrm{Sym}} (X) (\varepsilon_k \otimes \varepsilon_l) &=& \Phi_{\textrm{Fund}} (X) (\varepsilon_k) \otimes  \Phi_{\textrm{Fund}} (X) (\varepsilon_l) \, ,  \\
       \Phi_{\textrm{Adj}} (X) (\varepsilon_k \wedge \varepsilon_l) &=& \Phi_{\textrm{Fund}} (X) (\varepsilon_k) \wedge  \Phi_{\textrm{Fund}} (X) (\varepsilon_l) \, . 
\end{eqnarray}
Using these definitions, one easily computes the matrix representation of any element of $G$. 

As an example let us examine how this construction works for $n=4$. We use this material in the context of the moduli space of 4 instantons of $\mathrm{Sp}(1)$ in section \ref{subsectionHiggsBranchComputation}. We have to compute the images under $\Phi_{\textrm{Adj}}$ and $\Phi_{\textrm{Sym}}$ of the matrices $z$ and $P$ defined in (\ref{matrixz}) and (\ref{defP}) respectively. Using the images of individual basis vectors (see Table \ref{TableComputationAdj} and Table \ref{TableComputationSym}), one obtains 
\begin{equation}
\Phi_{\textrm{Adj}} (z) = \left(
\begin{array}{cccccc}
 1 & 0 & 0 & 0 & 0 & 0 \\
 0 & z_1 z_2 & 0 & 0 & 0 & 0 \\
 0 & 0 & \frac{z_1}{z_2} & 0 & 0 & 0 \\
 0 & 0 & 0 & \frac{z_2}{z_1} & 0 & 0 \\
 0 & 0 & 0 & 0 & \frac{1}{z_1 z_2} & 0 \\
 0 & 0 & 0 & 0 & 0 & 1 \\
\end{array}
\right)
\, , \qquad 
     \Phi_{\textrm{Sym}} (z) = \left(
\begin{array}{cccccccccc}
 z_1^2 & 0 & 0 & 0 & 0 & 0 & 0 & 0 & 0 & 0 \\
 0 & 1 & 0 & 0 & 0 & 0 & 0 & 0 & 0 & 0 \\
 0 & 0 & z_1 z_2 & 0 & 0 & 0 & 0 & 0 & 0 & 0 \\
 0 & 0 & 0 & \frac{z_1}{z_2} & 0 & 0 & 0 & 0 & 0 & 0 \\
 0 & 0 & 0 & 0 & \frac{1}{z_1^2} & 0 & 0 & 0 & 0 & 0 \\
 0 & 0 & 0 & 0 & 0 & \frac{z_2}{z_1} & 0 & 0 & 0 & 0 \\
 0 & 0 & 0 & 0 & 0 & 0 & \frac{1}{z_1 z_2} & 0 & 0 & 0 \\
 0 & 0 & 0 & 0 & 0 & 0 & 0 & z_2^2 & 0 & 0 \\
 0 & 0 & 0 & 0 & 0 & 0 & 0 & 0 & 1 & 0 \\
 0 & 0 & 0 & 0 & 0 & 0 & 0 & 0 & 0 & \frac{1}{z_2^2} \\
\end{array}
\right) \, , 
\end{equation}
\begin{equation}
   \Phi_{\textrm{Adj}} (P) = \left(
\begin{array}{cccccc}
 1 & 0 & 0 & 0 & 0 & 0 \\
 0 & 0 & 1 & 0 & 0 & 0 \\
 0 & 1 & 0 & 0 & 0 & 0 \\
 0 & 0 & 0 & 0 & 1 & 0 \\
 0 & 0 & 0 & 1 & 0 & 0 \\
 0 & 0 & 0 & 0 & 0 & -1 \\
\end{array}
\right)\, , \qquad 
 \Phi_{\textrm{Sym}} (P) = \left(
\begin{array}{cccccccccc}
 1 & 0 & 0 & 0 & 0 & 0 & 0 & 0 & 0 & 0 \\
 0 & 1 & 0 & 0 & 0 & 0 & 0 & 0 & 0 & 0 \\
 0 & 0 & 0 & 1 & 0 & 0 & 0 & 0 & 0 & 0 \\
 0 & 0 & 1 & 0 & 0 & 0 & 0 & 0 & 0 & 0 \\
 0 & 0 & 0 & 0 & 1 & 0 & 0 & 0 & 0 & 0 \\
 0 & 0 & 0 & 0 & 0 & 0 & 1 & 0 & 0 & 0 \\
 0 & 0 & 0 & 0 & 0 & 1 & 0 & 0 & 0 & 0 \\
 0 & 0 & 0 & 0 & 0 & 0 & 0 & 0 & 0 & 1 \\
 0 & 0 & 0 & 0 & 0 & 0 & 0 & 0 & 1 & 0 \\
 0 & 0 & 0 & 0 & 0 & 0 & 0 & 1 & 0 & 0 \\
\end{array}
\right) \, . 
\end{equation}
Similarly for $n=2$ the computation gives 
\begin{equation}
\Phi_{\textrm{Adj}} (z) = 1
\, , \qquad 
     \Phi_{\textrm{Sym}} (z) = \left(
\begin{array}{ccc}
 z_1^2 & 0 & 0 \\
 0 & 1 & 0 \\
 0 & 0 & \frac{1}{z_1^2} \\
\end{array}
\right) \, , 
\end{equation}
\begin{equation}
   \Phi_{\textrm{Adj}} (P) = -1 \, , \qquad 
 \Phi_{\textrm{Sym}} (P) = \left(
\begin{array}{ccc}
 0 & 0 & 1 \\
 0 & 1 & 0 \\
 1 & 0 & 0 \\
\end{array}
\right) \, . 
\end{equation}

\begin{table}[t]
    \centering
    \begin{tabular}{|c|c|c|}
    \hline 
    Vector & Image under $\Phi_{\textrm{Adj}} (z)$ & Image under $\Phi_{\textrm{Adj}} (P)$  \\  \hline  
    $\varepsilon_1 \wedge \varepsilon _2$ & $\varepsilon_1 \wedge \varepsilon _2$ & $\varepsilon_1 \wedge \varepsilon _2$     \\
    $\varepsilon_1 \wedge \varepsilon _3$ & $z_1z_2\varepsilon_1 \wedge \varepsilon _3$ & $\varepsilon_1 \wedge \varepsilon _4$     \\
    $\varepsilon_1 \wedge \varepsilon _4$ & $z_1z_2^{-1}\varepsilon_1 \wedge \varepsilon _4$ & $\varepsilon_1 \wedge \varepsilon _3$     \\
    $\varepsilon_2 \wedge \varepsilon _3$ & $z_1^{-1}z_2\varepsilon_2 \wedge \varepsilon _3$ & $\varepsilon_2 \wedge \varepsilon _4$     \\
    $\varepsilon_2 \wedge \varepsilon _4$ & $z_1^{-1}z_2^{-1}\varepsilon_2 \wedge \varepsilon _4$ & $\varepsilon_2 \wedge \varepsilon _3$     \\
    $\varepsilon_3 \wedge \varepsilon _4$ & $\varepsilon_3 \wedge \varepsilon _4$ & $-\varepsilon_3 \wedge \varepsilon _4$     \\
         \hline 
    \end{tabular}
    \caption{Detail of the computation of $\Phi_{\textrm{Adj}} (z)$ and $\Phi_{\textrm{Adj}} (P)$ in $\mathrm{O}(4)$.  }
    \label{TableComputationAdj}
\end{table}

\begin{table}[t]
    \centering
    \begin{tabular}{|c|c|c|}
    \hline 
    Vector & Image under $\Phi_{\textrm{Sym}} (z)$ & Image under $\Phi_{\textrm{Sym}} (P)$  \\  \hline  
    $\varepsilon_1 \otimes \varepsilon _1$ & $z_1^2\varepsilon_1 \otimes \varepsilon _1$ & $\varepsilon_1 \otimes \varepsilon _1$     \\
    $\varepsilon_1 \otimes \varepsilon _2$ & $\varepsilon_1 \otimes \varepsilon _2$ & $\varepsilon_1 \otimes \varepsilon _2$     \\
    $\varepsilon_1 \otimes \varepsilon _3$ & $z_1z_2\varepsilon_1 \otimes \varepsilon _3$ & $\varepsilon_1 \otimes \varepsilon _4$     \\
    $\varepsilon_1 \otimes \varepsilon _4$ & $z_1z_2^{-1} \varepsilon_1 \otimes \varepsilon _4$ & $\varepsilon_1 \otimes \varepsilon _3$     \\
    $\varepsilon_2 \otimes \varepsilon _2$ & $z_1^{-2}\varepsilon_2 \otimes \varepsilon _2$ & $\varepsilon_2 \otimes \varepsilon _2$     \\
    $\varepsilon_2 \otimes \varepsilon _3$ & $z_1^{-1}z_2\varepsilon_2 \otimes \varepsilon _3$ & $\varepsilon_2 \otimes \varepsilon _4$     \\
    $\varepsilon_2 \otimes \varepsilon _4$ & $z_1^{-1}z_2^{-1}\varepsilon_2 \otimes \varepsilon _4$ & $\varepsilon_2 \otimes \varepsilon _3$     \\
    $\varepsilon_3 \otimes \varepsilon _3$ & $z_2^2\varepsilon_3 \otimes \varepsilon _3$ & $\varepsilon_4 \otimes \varepsilon _4$     \\
    $\varepsilon_3 \otimes \varepsilon _4$ & $\varepsilon_3 \otimes \varepsilon _4$ & $\varepsilon_3 \otimes \varepsilon _4$     \\
    $\varepsilon_4 \otimes \varepsilon _4$ & $z_2^{-2}\varepsilon_3 \otimes \varepsilon _4$ & $\varepsilon_3 \otimes \varepsilon _3$     \\
         \hline 
    \end{tabular}
    \caption{Detail of the computation of $\Phi_{\textrm{Sym}} (z)$ and $\Phi_{\textrm{Sym}} (P)$ in $\mathrm{O}(4)$.  }
    \label{TableComputationSym}
\end{table}

\section{Proofs}

\subsection{The Jacobian}
\label{appendixJacobian}

In this appendix, we review the standard proof of (\ref{detDiff1}), based on the textbook \cite{knapp2013lie}. Recall that for $X \in G$ and $\mathbf{x} \in \mathfrak{g}$, we have 
\begin{equation}
   e^{\mathrm{Ad}(X) \mathbf{x}} = X e^{\mathbf{x}} X^{-1} \, . 
\end{equation}
So we can compute, with the notations of section \ref{sectionWeyl}, 
\begin{eqnarray}
        \psi \left( y e^{\epsilon \mathbf{y}} , z \right) &=& y e^{\epsilon \mathbf{y}} z e^{-\epsilon \mathbf{y}} y^{-1} \\
        &=& yzy^{-1} \left(y z^{-1} \right) e^{\epsilon \mathbf{y}} \left(y z^{-1} \right)^{-1} \left( y e^{-\epsilon \mathbf{y}} y^{-1} \right) \\
        &=& \psi \left( y , z \right) e^{\epsilon \mathrm{Ad}(y z^{-1}) \mathbf{y}   } e^{-\epsilon \mathrm{Ad}(y ) \mathbf{y}   } \\
        &=& \psi \left( y , z \right) e^{\epsilon \left( \mathrm{Ad}(y)  (\mathrm{Ad}(z^{-1}) -1\right) \mathbf{y}   } + O(\epsilon^2) \, ,
\end{eqnarray}
which means that 
\begin{equation}
   \mathrm{d}  \psi_{(y,z)} (\mathbf{y} , 0) =  \mathrm{Ad}(y)  \left(\mathrm{Ad}(z^{-1}) -1\right) \mathbf{y} \, . 
\end{equation}
Similarly, we compute $\psi \left( y  , z e^{\epsilon \mathbf{z}}\right) = y  z e^{\epsilon \mathbf{z}} y^{-1} = \psi \left( y  , z \right) e^{\epsilon  \mathrm{Ad}(y)\mathbf{z} } $ so that $ \mathrm{d}  \psi_{(y,z)} (0 , \mathbf{z}) =  \mathrm{Ad}(y)  \mathbf{z} $. By linearity, we then have 
\begin{equation}
   \mathrm{d}  \psi_{(y,z)} (\mathbf{y} , \mathbf{z}) = \mathrm{Ad}(y) \left(  (\mathrm{Ad}(z^{-1}) -1) \mathbf{y} + \mathbf{z} \right) \, . 
\end{equation}
On the Lie algebra $\mathfrak{t}$ of the maximal torus and on its complement $\mathfrak{t}^{\perp}$ the differential restricts to 
\begin{eqnarray}
        \mathrm{d}  \psi_{(y,z)} |_{\mathfrak{t}} &=& \mathrm{Ad}(y)    \\
        \mathrm{d}  \psi_{(y,z)} |_{\mathfrak{t}^{\perp}} &=& \mathrm{Ad}(y)    (\mathrm{Ad}(z^{-1}) -1) \, . 
\end{eqnarray}
Noting that $\det \mathrm{Ad}(y) = 1$ we obtain the first part of (\ref{detDiff1}). To compute the determinant, we recall the root-space decomposition 
\begin{equation}
    \mathfrak{g} = \mathfrak{t} \oplus \bigoplus\limits_{\alpha \in \Delta (\mathfrak{g})} \mathfrak{g}_{\alpha} \, ,
\end{equation}
where by definition the $\mathfrak{g}_{\alpha}$ are the eigenspaces of $\mathrm{Ad}(z)$ with eigenvalue $z^{\alpha}$. Thus we obtain the second part of (\ref{detDiff1}). 

\subsection{Equivalence between two Measures}
\label{appendixMeasures}

This subsection is devoted to the proof of (\ref{equalityMeasures}). We start with a useful lemma: 
\begin{equation}
\label{lemma}
    \prod\limits_{\alpha} (1-z^\alpha) = \sum\limits_{w \in W} \prod\limits_{\alpha>0} (1-z^{w(\alpha)}) \, . 
\end{equation}
We prove this Lemma as follows:\footnote{We thank Friedrich Knop for helping us in writing this proof. }
\begin{eqnarray*}
    \prod\limits_{\alpha>0} (1-z^{w(\alpha)}) &=& \prod\limits_{\substack{\alpha>0,\\ w^{-1} (\alpha)>0}} (1-z^{\alpha}) \prod\limits_{\substack{\alpha<0,\\ w^{-1} (\alpha)>0}} (1-z^{\alpha}) \\
    &=& \prod\limits_{\substack{\alpha>0,\\ w^{-1} (\alpha)>0}} (1-z^{\alpha}) \prod\limits_{\substack{\alpha>0,\\ w^{-1} (\alpha)<0}} (1-z^{-\alpha}) \\
    &=& \prod\limits_{\substack{\alpha>0,\\ w^{-1} (\alpha)>0}} (1-z^{\alpha}) \prod\limits_{\substack{\alpha>0,\\ w^{-1} (\alpha)<0}} (-z^{-\alpha})(1-z^{\alpha}) \\
    &=& (-1)^{\ell (w)} z^{w(\rho) - \rho} \prod\limits_{\substack{\alpha>0,\\ w^{-1} (\alpha)>0}} (1-z^{\alpha}) \prod\limits_{\substack{\alpha>0,\\ w^{-1} (\alpha)<0}} (1-z^{\alpha}) \\
    &=& (-1)^{\ell (w)} z^{w(\rho) - \rho} \prod\limits_{\alpha>0 } (1-z^{\alpha}) \, . 
\end{eqnarray*}
Here we have used the length $\ell (w)$ of the Weyl group element and the Weyl vector $\rho$, equal to the half-sum of the positive roots, noticing that
\begin{equation}
    \rho - w(\rho) = \frac{1}{2} \sum\limits_{\alpha>0} \alpha  - \frac{1}{2} \sum\limits_{w^{-1}(\alpha)>0}  \alpha  = \sum\limits_{\substack{\alpha>0,\\ w^{-1} (\alpha)<0}} \alpha \, . 
\end{equation}
Then using the Weyl denominator formula,  
\begin{eqnarray*}
    \sum\limits_{w \in W} \prod\limits_{\alpha>0} (1-z^{w(\alpha)}) &=& \left( \sum\limits_{w \in W} (-1)^{\ell (w)} z^{w(\rho) - \rho} \right) \prod\limits_{\alpha>0 } (1-z^{\alpha}) \\ 
    &=& \prod\limits_{\alpha>0 } (1-z^{-\alpha}) \prod\limits_{\alpha>0 } (1-z^{\alpha}) \\ 
    &=& \prod\limits_{\alpha }  (1-z^{\alpha}) \, . 
\end{eqnarray*}

Now using the lemma (\ref{lemma}), we show that for any function $f(z)$ that is invariant under the Weyl group, 
\begin{eqnarray*}
\int_T \d \mu_G (z) f(z) 
    &=& \left( \prod\limits_{l=1}^r \oint_{|z_l|=1} \frac{\d z_l}{z_l} \right) \prod\limits_{\alpha>0} (1-z^{-\alpha}) f(z)\\
    &=& \frac{1}{|W|} \sum\limits_{w \in W}\left( \prod\limits_{l=1}^r \oint_{|z_l|=1} \frac{\d z_l}{z_l} \right) \prod\limits_{\alpha>0} (1-z^{-\alpha}) f(w(z)) \\
    &=& \frac{1}{|W|} \sum\limits_{w \in W}\left( \prod\limits_{l=1}^r \oint_{|z_l|=1} \frac{\d z_l}{z_l} \right) \prod\limits_{\alpha>0} (1-z^{-w(\alpha)}) f(z) \\
    &=& \frac{1}{|W|} \left( \prod\limits_{l=1}^r \oint_{|z_l|=1} \frac{\d z_l}{z_l} \right) \prod\limits_{\alpha} (1-z^{-\alpha}) f(z) \\
    &=& \int_T \d \tilde{\mu}_G (z) f(z) \, . 
\end{eqnarray*}
This concludes the proof of (\ref{equalityMeasures}).

\section{\texorpdfstring{Hilbert Series for $k$ $\mathrm{Sp}(N)$ instantons}{}}
\label{appendix:hilbertseries}

In this appendix we report some results obtained for the unrefined Hilbert series for $k=4$ and $N=2,3$ and for $k=6$ and $N=1,2$. 

\paragraph{\texorpdfstring{Four $\mathrm{Sp}(2)$ instantons}{}}
\begin{align*}
g_{2,4}(t) &= \frac{1}{(1-t)^{24} (1+t)^{12} \left(1+t^2\right)^6 \left(1+t+t^2\right)^9 \left(1+t+t^2+t^3+t^4\right)^5}\times \\
& (1+4 t+17 t^{2}+63 t^{3}+222 t^{4}+714 t^{5}+2163 t^{6}+6147 t^{7}+16574 t^{8}+42337 t^{9}+\\
& 102823 t^{10}+237512 t^{11}+522911 t^{12}+1098007 t^{13}+2201815 t^{14}+4219555 t^{15}+\\
& 7735656 t^{16}+13576835t^{17}+22832527 t^{18}+36819853 t^{19}+56980025 t^{20}+84678080 t^{21}+\\
& 120927442 t^{22}+166051921 t^{23}+219370384 t^{24}+278958448 t^{25}+341603953 t^{26}+\\
& 402982089 t^{27}+458104174 t^{28}+501948075 t^{29}+530208657t^{30}+539970736 t^{31} +\\
& + \ \textrm{palindrome} \ + t^{62})\, .
\end{align*}

\paragraph{\texorpdfstring{Four $\mathrm{Sp}(3)$ instantons}{}}

\begin{align*}
g_{2,6}(t) & = \frac{1}{(1-t)^{32} (1+t)^{18} \left(1+t^2\right)^8 \left(1+t+t^2\right)^{13} \left(1+t+t^2+t^3+t^4\right)^7}\\
& (1+8 t+50 t^2+265 t^3+1263 t^4+5519 t^5+22506 t^6+86249 t^7+312486 t^8+\\
& 1073961 t^9+3510000 t^{10}+10926327 t^{11}+32436958 t^{12}+91926975 t^{13}+248940988 t^{14}+\\
& 644771065 t^{15}+1598791859 t^{16}+3799215663t^{17}+8660987282 t^{18}+18961709262 t^{19}+\\
& 39910959120 t^{20}+80848915242 t^{21}+157788448901 t^{22}+296982261477 t^{23}+\\
& 539579851501 t^{24}+947205668718 t^{25}+1607926370644 t^{26}+2641600371012 t^{27}+\\
& 4203090504941t^{28}+6481397831797 t^{29}+9692621192295 t^{30}+14064947226086 t^{31}+\\
& 19814776064818 t^{32}+27114609659596 t^{33}+36055332342145 t^{34}+46607643703755 t^{35}+\\
& 58589165281807 t^{36}+71644556206784 t^{37}+85245376694700t^{38}+98714201635720 t^{39}+\\
& 111273795336187 t^{40}+122117586312437 t^{41}+130493171949721 t^{42}+\\
& 135787171120594 t^{43}+137598361834214 t^{44} + \ \textrm{palindrome} \  + t^{88})\, .
\end{align*}

\paragraph{\texorpdfstring{Six $\mathrm{Sp}(1)$ instantons}{}} 

\begin{eqnarray*}
     g_{1,6}(t) &=& \frac{1}{(1-t)^{24} (1+t)^{12} \left(1+t^2\right)^4
   \left(1-t+t^2\right)^4 \left(1+t+t^2\right)^8} \times \\
   & &\frac{1}{\left(1+t+t^2+t^3+t^4\right)^5
   \left(1+t+t^2+t^3+t^4+t^5+t^6\right)^3} \times \\
   & & (1+2 t+5 t^{2}+14 t^{3}+36 t^{4}+83 t^{5}+193
   t^{6}+422 t^{7}+892 t^{8}+1821 t^{9}\\
   & &+3620 t^{10}+6955
   t^{11}+13017 t^{12}+23649 t^{13}+41856 t^{14}+72130
   t^{15}+121233 t^{16}\\
   & &+198686 t^{17}+317998 t^{18}+496951
   t^{19}+759026 t^{20}+1133300 t^{21}+1655290 t^{22}\\
   & &+2365512
   t^{23}+3309485 t^{24}+4533761 t^{25}+6084418 t^{26}+8000798
   t^{27}\\
   & &+10312362 t^{28}+13030773 t^{29}+16147551 t^{30}+19625914
   t^{31}+23401717 t^{32}\\
   & &+27378910 t^{33}+31435436 t^{34}+35423981
   t^{35}+39184907 t^{36}+42550833 t^{37}\\
   & &+45364374 t^{38}+47484587
   t^{39}+48803727 t^{40}+49250804 t^{41} + \textrm{palindrome} +t^{82})  \, .  
\end{eqnarray*}

\paragraph{\texorpdfstring{Six $\mathrm{Sp}(2)$ instantons}{}} 

\begin{eqnarray*}
    g_{2,6}(t) &=&\frac{1}{(1-t)^{36} (1+t)^{18} \left(1+t^2\right)^9
   \left(1-t+t^2\right)^6 \left(1+t+t^2\right)^{12} \left(1+t+t^2+t^3+t^4\right)^7} \times \\
   & & \frac{1}{\left(1+t+t^2+t^3+t^4+t^5+t^6\right)^5} \times ( 1+2 t+13 t^2+42 t^3+152 t^4+481 t^5+1512 t^6 \\
   & & +4446 t^7+12793 t^8+35315 t^9+94958
   t^{10}+247472 t^{11}+628523 t^{12}+1553689 t^{13} \\
   & & +3747474 t^{14}+8817761
   t^{15}+20266024 t^{16}+45503832 t^{17}+99887900 t^{18} \\
   & & +214429359 t^{19}+450382031
   t^{20}+925834317 t^{21}+1863431217 t^{22}+3673264612 t^{23} \\
   & & +7094132051
   t^{24}+13427139508 t^{25}+24913929895 t^{26}+45331673421 t^{27} \\
   & & +80908050053
   t^{28}+141688695934 t^{29}+243531924471 t^{30}+410934694795 t^{31} \\
   & & +680932502910
   t^{32}+1108316475589 t^{33}+1772410187153 t^{34}+2785568842110 t^{35} \\
   & & +4303469079760
   t^{36}+6537034739643 t^{37}+9765609758659 t^{38}+14350590646180 t^{39} \\
   & & +20748373662928
   t^{40}+29520957244303 t^{41}+41342201351789 t^{42} \\
   & & +56997316832910
   t^{43}+77373123134561 t^{44}+103436586844630 t^{45} \\
   & & +136199767790649
   t^{46}+176669959408685 t^{47}+225785270073232 t^{48} \\
   & & +284337245351281
   t^{49}+352884238504866 t^{50}+431660811482260 t^{51} \\
   & & +520490499984005
   t^{52}+618710151378632 t^{53}+725115027370307 t^{54} \\
   & & +837933042764956
   t^{55}+954835624577211 t^{56}+1072989695329237 t^{57} \\
   & & +1189152539636575
   t^{58}+1299806735524741 t^{59}+1401328836092757 t^{60} \\
   & & +1490180890015773
   t^{61}+1563111493138243 t^{62}+1617350369495078 t^{63} \\
   & & +1650780865697407
   t^{64}+1662075183972652 t^{65}+ \textrm{palindrome} +t^{130}) \, . 
\end{eqnarray*}

\bibliographystyle{JHEP}
\bibliography{bibli.bib}

\end{document}